\pdfoutput=1
\documentclass[11pt]{article}

\usepackage{acl}

\usepackage{times}
\usepackage{latexsym}

\usepackage[T1]{fontenc}
\usepackage[utf8]{inputenc}

\usepackage{microtype}

\usepackage{multirow}
\usepackage{booktabs}
\usepackage{graphicx}
\usepackage{amsmath}

\title{Defense of Adversarial Ranking Attack in Text Retrieval: \\ Benchmark and Baseline via Detection}

\author{Xuanang Chen$^{\rm 1,2}$
        \quad Ben He$^{\rm 1,2}$
        \quad Le Sun$^{\rm 2}$
        \quad Yingfei Sun$^{\rm 1}$ \\
        $^{\rm 1}$University of Chinese Academy of Sciences, Beijing, China \\
        $^{\rm 2}$Institute of Software, Chinese Academy of Sciences, Beijing, China \\
        \texttt{chenxuanang19@mails.ucas.ac.cn, benhe@ucas.ac.cn}\\
        \texttt{sunle@iscas.ac.cn, yfsun@ucas.ac.cn}}
        
\begin{document}
\maketitle
\begin{abstract}
Neural ranking models (NRMs) have undergone significant development and have become integral components of information retrieval (IR) systems. 
Unfortunately, recent research has unveiled the vulnerability of NRMs to adversarial document manipulations, potentially exploited by malicious search engine optimization practitioners.
While progress in adversarial attack strategies aids in identifying the potential weaknesses of NRMs before their deployment, the defensive measures against such attacks, like the detection of adversarial documents, remain inadequately explored.
To mitigate this gap, this paper establishes a benchmark dataset to facilitate the investigation of adversarial ranking defense and introduces two types of detection tasks for adversarial documents.
A comprehensive investigation of the performance of several detection baselines is conducted, which involve examining the spamicity, perplexity, and linguistic acceptability, and utilizing supervised classifiers. 
Experimental results demonstrate that a supervised classifier can effectively mitigate known attacks, but it performs poorly against unseen attacks. 
Furthermore, such classifier should avoid using query text to prevent learning the classification on relevance, as it might lead to the inadvertent discarding of relevant documents.
\end{abstract}

\section{Introduction}
In information retrieval (IR) systems, neural ranking models (NRMs) offer substantial performance improvements in the re-ranking stage by finer-grained interactions between queries and documents, particularly those that utilize pre-trained language models (PLMs)~\cite{DBLP:journals/corr/abs-1901-04085,DBLP:series/synthesis/2021LinNY}.
However, besides effectiveness, NRMs have been shown to inherit adversarial vulnerabilities of general neural networks~\cite{DBLP:journals/corr/SzegedyZSBEGF13}, which raises legitimate concerns about the robustness and trustworthiness of neural IR systems, and receives increasing attention from the research community~\cite{DBLP:journals/tois/WuZGFC23}.
Therefore, there have been quite a few initial studies on the adversarial attacks by adding small deliberate perturbations in the input documents to cause a catastrophic ranking disorder in the outcome of NRMs~\cite{DBLP:journals/corr/abs-2204-01321,DBLP:conf/ccs/LiuKTSSWLL22,DBLP:conf/ictir/WangLA22,song-etal-2022-trattack,DBLP:conf/acl/ChenHY0S23,DBLP:conf/sigir/Liu0GR0FC23}, which aim to investigate and identify the vulnerability of NRMs before deploying them in real-world applications.

\begin{figure}
    \centering
    \includegraphics[width=\linewidth]{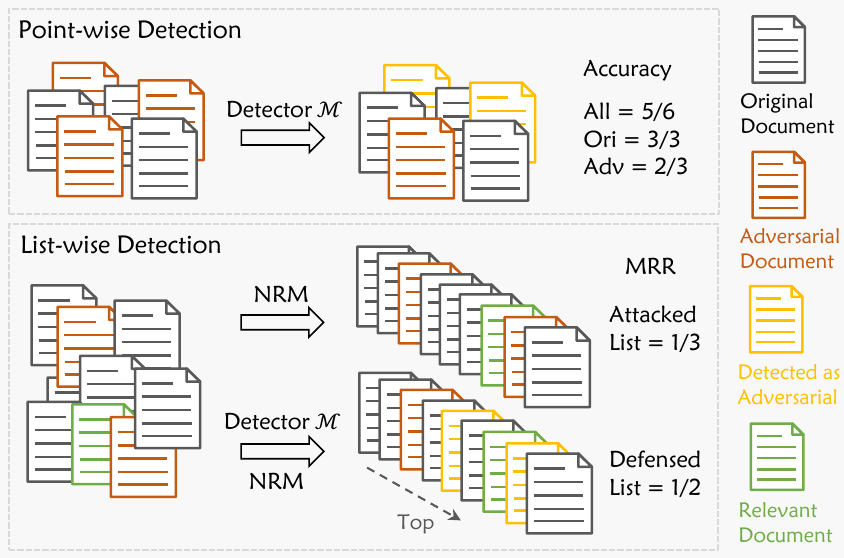}
    \caption{Two types of detection tasks against adversarial documents. The point-wise detection primarily emphasizes the overall accuracy of the detection, while the list-wise detection further considers the ranking quality (e.g., MRR metric) of the final ranking list.}
    \label{fig.detection}
\end{figure}

Overall, current works mainly focus on the attack side, and take efforts to design general document manipulation frameworks that can provide better attack effectiveness and higher success rate, such as from word substitution~\cite{DBLP:journals/corr/abs-2008-02197,DBLP:journals/corr/abs-2204-01321} to trigger injection~\cite{DBLP:conf/ccs/LiuKTSSWLL22,DBLP:conf/acl/ChenHY0S23}, and from query-specific to topic-oriented attack~\cite{DBLP:conf/sigir/Liu0GR0FC23}.
However, the research on the defense side to combat such attacks is lacking, such as conducting a systematic investigation into the detection methods of adversarial documents.
While some basic tools, such as anti-spam methods and online grammar checkers, have been employed to assess the naturalness of adversarial documents~\cite{DBLP:journals/corr/abs-2204-01321,DBLP:conf/ccs/LiuKTSSWLL22,DBLP:conf/sigir/Liu0GR0FC23}, they fall short of providing adequate support for developing robust countermeasures to build a trustworthy ranking system against adversarial attacks.

In light of this gap, this paper aims to introduce a benchmark dataset to facilitate studies focused on the defense of adversarial ranking attacks in text retrieval.
Specifically, we utilize the widely-used MS MARCO passage dataset~\cite{DBLP:conf/nips/NguyenRSGTMD16} as the primary data source, from which we gather different pairs of query and original document. 
To generate corresponding adversarial documents, we employ three novel attack methods, namely PRADA~\cite{DBLP:journals/corr/abs-2204-01321}, PAT~\cite{DBLP:conf/ccs/LiuKTSSWLL22} and IDEM~\cite{DBLP:conf/acl/ChenHY0S23}. 
As a result, this synthetic benchmark dataset contains about 144K, 10K, 10K adversarial examples in its train, valid and test sets, respectively.
Besides, as depicted in Figure~\ref{fig.detection}, we introduce two distinct detection tasks tailored for adversarial documents.
The point-wise detection task solely focuses on evaluating the detection accuracy, while the list-wise detection task incorporates the simulation of the text retrieval process to assess the ranking quality.

After that, we investigate the effectiveness of several detection methods, including unsupervised detectors based on spamicity~\cite{DBLP:conf/sdm/ZhouPT08}, perplexity~\cite{DBLP:journals/coling/BrownPPLM92} or linguistic acceptability~\cite{DBLP:journals/tacl/WarstadtSB19}, and supervised classification detectors based on BERT~\cite{devlin-etal-2019-bert} and RoBERTa~\cite{DBLP:journals/corr/abs-1907-11692} models.
Based on our extensive experiments, we observe that the supervised classification detector performs exceptionally well in the point-wise detection task when trained on all types of adversarial documents. However, this level of accuracy cannot be sustained when the detector encounters unknown types of adversarial documents during testing.
Furthermore, we notice that while supervised classification training with both the query and document text can improve accuracy, incorporating query text into the training process could result in a relevance-aware detector. 
This relevance-awareness increases the likelihood of misidentifying a relevant document as an adversarial one, which, in turn, negatively impacts the performance of the detector in the list-wise detection task.

Our contributions are three-fold: 
1) This paper represents the first-ever investigation into the defense of adversarial ranking attacks, and a benchmark dataset\footnote{This dataset is available at \url{https://github.com/cxa-unique/DARA} for future research.} is proposed to support further research in this area.
2) Two kinds of detection tasks, namely point-wise and list-wise detection, are introduced to standardize evaluation processes of the efficacy of adversarial ranking detection methods.
3) Our research comprehensively examines several detection methods for their effectiveness in defense, yielding intriguing results that can serve as valuable clues for future studies.

\section{Benchmark}
To support the investigation on the defense of adversarial ranking attacks in text retrieval, we attempt to construct a synthetic dataset that contains a series of original and adversarial document pairs.
This synthetic dataset is on top of the widely-used MS MARCO passage dataset~\citep{DBLP:conf/nips/NguyenRSGTMD16} and several representative attack methods. 

\subsection{Attack Method}\label{sec.attack.methods}
The adversarial documents are produced by three recently proposed black-box attack methods, which employ different perturbations,such as word substitutions and trigger insertions to manipulate the content of the documents.

\textbf{PRADA}~\citep{DBLP:journals/corr/abs-2204-01321} launch attacks on word level, it first finds important words (i.e., sub-word tokens) in the target document according to the gradient magnitude~\cite{DBLP:journals/eaai/XuD20}, and then greedily replaces them with the synonyms found in a perturbed word embedding space via PGD~\cite{DBLP:conf/iclr/MadryMSTV18}.

\textbf{PAT}~\citep{DBLP:conf/ccs/LiuKTSSWLL22} generates and adds several trigger tokens at the beginning of the target document, its search objective is equipped with semantic and fluency constraints using the pre-trained BERT model~\citep{devlin-etal-2019-bert} in addition to the ranking-incentivized objective.

\textbf{IDEM}~\citep{DBLP:conf/acl/ChenHY0S23} instructs the BART model~\citep{lewis-etal-2020-bart} to generate a series of connection sentences between the query and the target document, inserts connection sentences to produce adversarial documents, and finds the most adversarial and coherent one by a position-wise merging mechanism.

Akin to PAT~\citep{DBLP:conf/ccs/LiuKTSSWLL22} and IDEM~\citep{DBLP:conf/acl/ChenHY0S23}, we employ the ‘msmarco-MiniLM-L-12-v2’ model publicly available at Sentence-Transformers~\cite{reimers-gurevych-2019-sentence} as the representative victim NRM in this work, due to its highly effective ranking performance on the MS MARCO Dev set.
All these three methods rely on a surrogate NRM to evaluate the attack effect or determine the attack direction, we follows the instructions provided in IDEM~\citep{DBLP:conf/acl/ChenHY0S23} for the training of this surrogate model.

\subsection{Dataset Collection}\label{sec.data}
Akin to recent research efforts focused on designing adversarial ranking attacks~\cite{DBLP:journals/corr/abs-2204-01321,DBLP:conf/ccs/LiuKTSSWLL22,DBLP:conf/sigir/Liu0GR0FC23,DBLP:conf/acl/ChenHY0S23}, we chose to create and gather adversarial data on top of the popular MS MARCO passage dataset~\citep{DBLP:conf/nips/NguyenRSGTMD16}.
This paragraph-level dataset contains approximately 8.8 million passages as the corpus, along with about 503 thousand train queries with relevance labels and 6,980 Dev queries for evaluation.
To construct our dataset, we randomly selected 50 thousand train queries to form our Train set, and evenly divided the Dev set of MS MARCO into two parts, designating them as our Valid and Test sets for evaluation purposes.

\textbf{Target documents.}
Considering a great reproducibility of the BM25 model using the Anserini toolkit~\cite{DBLP:journals/jdiq/YangFL18}, we chose to sample target documents from the top-1000 BM25 candidates for all queries.
We threw away a query in the Test set as it has less than 50 BM25 candidates, so the Valid and Test sets contain 3,490 and 3,489 queries, respectively.
The victim NRM mentioned in Section~\ref{sec.attack.methods} would re-rank these BM25 candidates to produce the final re-ranking list, from which the target documents were sampled.
Specifically, for each query and each attack method, one target document ranked between [51, 1000] was randomly selected, and target documents for each attack method on the same query are different.

\textbf{Document manipulations.}
After obtaining the target documents, we carry out the attacks as described in Section~\ref{sec.attack.methods} to modify the original text of the target documents to be adversarial ones.
As all three attack methods do not have 100\% attack success rate, we only keep the adversarial documents that can be ranked higher than their original versions by the victim NRM.
As summarized in Table~\ref{tab:data.summary}, the number of adversarial samples are smaller than the number of queries in Train (50,000), Valid (3,490) and Test (3,489) sets.
Considering the difference in experimental equipment like GPU, the output results of the victim NRM could be non-deterministic, especially the relevance scores in float type.
Thus, we choose to only release the text data in the form of ``\textit{query id, query text, document id, original document text, adversarial document text}'' with out any information of the relevance scores or ranks.
Note that, this dataset can be expanded by adding more adversarial examples from more recently proposed attack methods.

\begin{table}[t]
\centering
\resizebox{0.9\linewidth}{!}{
\begin{tabular}{l|ccc}
\toprule
  \multirow{2}{*}{\textbf{Attack Method}} & \multicolumn{3}{c}{\textbf{\# of Adversarial Samples}} \\
  & \textbf{Train} & \textbf{Valid} & \textbf{Test} \\
  \midrule
  PRADA & 46,372 & 3,267 & 3,255 \\
  PAT & 47,698 & 3,341 & 3,338 \\
  IDEM & 49,827 & 3,471 & 3,472 \\
  \midrule
  ALL & 143,897 & 10,079 & 10,065 \\

\bottomrule
\end{tabular}
}
\caption{Summary of the benchmark dataset. An adversarial sample consists of a query, a target original document and adversarial document. }
\label{tab:data.summary}
\end{table}

\subsection{Detection Task}\label{sec.task}
On top of the constructed adversarial dataset, we can carry out defense experiments.
In this work, we focus on the detection of the adversarial documents in the first-stage retrieval candidates (such as by BM25 model) before feeding them into the NRMs for re-ranking.
Thus, we need a detector to find as many adversarial documents as possible while wrongful killing the original document as little as possible. 
Herein, we introduce two detection task settings and their evaluation metrics.

\textbf{Point-wise detection.}
Given a single document (and a specific query if it is needed), the task is to detect whether it is an adversarial one.
As illustrated in Figure~\ref{fig.detection}, given a set of documents, a detector $\mathcal{M}$ is used to check all documents and identify adversarial ones.
In the generated dataset in Section~\ref{sec.data}, the original and adversarial documents are paired, that is, the number of original documents is equal to the number of adversarial documents.
Therefore, in this point-wise detection, we use the average \textit{Accuracy} of the classification on all documents (both the original and adversarial ones) as the main evaluation metric.

\textbf{List-wise detection.}
In practical applications like search engine, the number of adversarial documents is often not equal to the number of original documents.
In most cases, normally, the original documents shall account for the majority.
Therefore, a detector should be good at finding the adversarial documents from a series of original documents, and also be friendly to these original documents, especially the relevant documents.
As illustrated in Figure~\ref{fig.detection}, given a query and a set of top-\textit{k} candidate documents from a retrieval model like BM25, a detector $\mathcal{M}$ is used to examine all these candidates, and discard those documents that are much likely to be adversarial ones.
In other word, only the documents that are judged as original or normal are fed into a NRM to obtain a more accurate relevance score.
In this setting, we use the metrics that are originally used to reflect the ranking quality, like \textit{MRR@K}, to evaluate the performance of list-wise detection.

\subsection{Detection Baseline}\label{sec.baseline}
In this study, we explore two types of detectors. The first type consists of unsupervised models that rely on various quality characteristics of adversarial documents, such as spamicity (OSD), perplexity (PPL), or linguistic acceptability (LA). The second type involves supervised classifiers based on pre-trained
language models (PLMs), such as BERT.

\textbf{Spamicity-based detector}.
In previous research, a utility-based term spamicity method known as OSD (Online Spam Detection;~\citealp{DBLP:conf/sdm/ZhouPT08,zhou2009osd}) has been utilized to identify adversarial examples. 
However, its previous primary function has been to evaluate the naturalness of adversarial documents~\cite{DBLP:journals/corr/abs-2204-01321,DBLP:conf/ccs/LiuKTSSWLL22}. 
OSD relies mainly on TF-IDF~\cite{DBLP:books/aw/Baeza-YatesR99} features and has been validated by the Microsoft adCenter~\cite{DBLP:conf/ccs/LiuKTSSWLL22}. 
Hence, we incorporate it into our study as a baseline detector. 
A document is filtered out if its OSD score surpasses a specific threshold.

\textbf{Perplexity-based detector}. 
As demonstrated in earlier research, adversarial perturbations applied to original documents can significantly impact the semantic fluency of their content~\cite{DBLP:conf/ccs/LiuKTSSWLL22,DBLP:conf/acl/ChenHY0S23}. 
Akin to~\citet{song-etal-2020-adversarial}, we can employ a perplexity-based strategy to counter ranking attacks. 
This strategy involves leveraging a pre-trained language model (PLM) to assess the perplexity of documents, where higher perplexity values indicate less fluent text. 
Consequently, any document surpassing a certain perplexity threshold is filtered out from consideration.

\textbf{Linguistic acceptability-based detector}.
Adversarially generated or modified documents often exhibit grammatical inconsistencies or lack context coherence~\cite{DBLP:journals/corr/abs-2302-05892,DBLP:conf/acl/ChenHY0S23}. Following the approach used by~\citet{DBLP:conf/acl/ChenHY0S23}, we employ a RoBERTa-based classification model\footnote{\url{https://huggingface.co/textattack/roberta-base-CoLA}} trained on the Corpus of Linguistic Acceptability dataset (CoLA;~\citealp{DBLP:journals/tacl/WarstadtSB19}) to determine the grammaticality of the document text.
Any document deemed to have poor linguistic acceptability is subsequently discarded.

\textbf{Supervised learning-based detector.}
As mentioned earlier, the OSD-based, PPL-based, and LA-based detectors lack knowledge of the adversarial documents, potentially leading to sub-optimal performance. 
Following recent advancements in detecting AI-generated text~\cite{DBLP:journals/corr/abs-2301-07597}, a deep classifier based on PLMs emerges as a strong candidate for addressing this kind of text classification problem. 
Consequently, we opt to fine-tune the BERT and RoBERTa models using the original and adversarial document pairs present in the Train set of the generated dataset. 
There are two versions of the classifier, depending on whether the query text is utilized or not. 
Intuitively, incorporating the query text can make it easier to determine if a single document is adversarial. 
However, as shown later in Section~\ref{sec.res.list}, using query text could increase the likelihood of discarding relevant documents, thereby potentially leading to failures in the list-wise detection task.

\section{Experiments}
\subsection{Experimental Setup}
\textbf{Instructions for our benchmark dataset.}
As described in Section~\ref{sec.data}, our dataset is in the format of ``\textit{query id, query text, document id, original document text, adversarial document text}''.
Before experimenting on this dataset, you need to prepare initial ranking lists containing the target documents used in this dataset, but the target victim NRM can be any one you want to attack.
Besides, you may observe that some adversarial documents cannot be ranked higher on your individual devices, even when using the same victim NRM employed in our study. 
It is highly probable that these failed adversarial documents were generated by PRADA, as their relevance score gains could be minimal. 
Additionally, our dataset is divided into different segments based on the attack method used, which enables us to perform cross-attack detection and assess the detector's ability to generalize across various types of adversarial attacks.
For example, the detector established on the PAT's adversarial documents can be tested on other single-attack set (such as PRADA in Table~\ref{tab:data.summary}) or to the whole-attack set (namely, ALL in Table~\ref{tab:data.summary}).

\textbf{Settings of detection tasks.}
As discussed in Section~\ref{sec.task}, our experimentation involves two types of detection tasks aimed at identifying adversarial documents. 
The point-wise detection primarily emphasizes the accuracy of detection, treating both original and adversarial documents equally. In contrast, the list-wise detection goes a step further by considering how adversarial documents affect the overall ranking quality.
In both detection tasks, given a document $d_{i}$ and a query $q_{i}$ (if necessary), the detector $\mathcal{M}$ assigns a judgement score $J_{\mathcal{M}}(d_{i})$ to this document, which reflects the likelihood of the document being an adversarial one. 
A score thread $\delta$ is typically established to determine the boundary between original and adversarial documents.
This threshold can be adjusted to achieve optimal detection performance on the corresponding sample set, such as the Train or Valid sets.
In the list-wise detection, we detect the top-1000 BM25 candidates and directly remove the adversarial ones, which may result in a final ranking list with less than 1000 documents.
Actually, in real-world systems, adversarial documents can influence the final ranking outcomes only when they are elevated to the top candidate set by the retrieval models. 
However, for simplicity, we adopt a streamlined approach~\cite{DBLP:conf/ccs/LiuKTSSWLL22,DBLP:conf/acl/ChenHY0S23}, where we solely focus on examining the impact of attacks on the re-ranking process, thereby avoiding the complexities of repetitive indexing.

\textbf{Details of detection baselines.}
As mentioned in Section~\ref{sec.baseline}, we conduct experiments using two types of detection methods. The unsupervised detector lacks any knowledge about the adversarial documents, whereas the supervised detector has been trained using the adversarial documents.

\textit{OSD-based detector} computes a term spamicity score between 0 and 1 to indicate the probability of the document being term spam~\cite{DBLP:conf/sdm/ZhouPT08}, and discards those documents with OSD values surpassing a pre-defined threshold $\delta_{osd}$.
The threshold $\delta_{osd}$ is adjusted between the smallest and largest OSD values among the given document set with a step of 0.01, to get a best accuracy on the Train set in the point-wise detection task and to get a best MRR on the Valid set in the list-wise detection task.
We adopt the OSD implementation released in PAT~\cite{DBLP:conf/ccs/LiuKTSSWLL22}.

\textit{PPL-based detector} builds upon the GPT-2~\cite{radford2019language} model in this study, it assigns perplexity values greater than zero to documents and filters out those with perplexity values exceeding a pre-defined threshold $\delta_{ppl}$.
This threshold is searched within the interval encompassing the smallest and largest PPL values in the given document set with a step of 0.1, to obtain the optimal accuracy on the Train set in the point-wise detection task and the best MRR on the Valid set in the list-wise detection task.

\textit{LA-based detector} utilizes the RoBERTa-based CoLA model as mentioned in Section~\ref{sec.baseline}, it generates linguistic acceptability scores ranging from 0 to 1 (logits after the Softmax function), and any document with an acceptability score lower than a specific threshold $\delta_{la}$ is filtered out
To achieve the optimal accuracy on the Train set for the point-wise detection task and the best MRR on the Valid set for the list-wise detection task, we search for threshold $\delta_{la}$ within the range of the smallest and largest LA values in the given document set, using a step size of 0.01.

\textit{SL-based detector} employs pre-trained language models (PLMs), such as BERT~\cite{devlin-etal-2019-bert} and RoBERTa~\cite{DBLP:journals/corr/abs-1907-11692}, as its backbone. 
It is fine-tuned under the supervision from Train set, which consists of both original (labeled as 0) and adversarial (labeled as 1) samples, using a learning rate of 2e-5 and a batch size of 32 with the cross-entropy loss. 
This type of classifier typically employs a Softmax function and returns the label with the higher probability, setting the boundary at 0.5. 
In the point-wise detection, the detector for single-attack (whole-attack) is fine-tuned on the single-attack (whole-attack) Train set for eight (four) epochs and selected every 1K (2K) training steps according to the average Accuracy achieved on the corresponding Valid set.
In the list-wise detection, we directly use the detector trained on the whole-attack Train set in the point-wise detection, so the ranking list for each test query contains at most three distinct adversarial documents.
Different to point-wise detection, we adjust the classification boundary (denoted as $\delta_{cls}$) with a step of 0.01 for better performance in list-wise detection.
Additionally, detectors designed to handle a single piece of document text are denoted as `w/o query', while those capable of receiving both the query and document texts are denoted as `w query'.

\begin{table*}[t]
\centering
\resizebox{\textwidth}{!}{
\begin{tabular}{ll|cc|cc|cc|cc}
\toprule
  \textbf{Tuning /} & \multirow{2}{*}{\textbf{Detector}} & \multicolumn{2}{c|}{\textbf{Test}$_{\text{PRADA}}$} & \multicolumn{2}{c|}{\textbf{Test$_{\text{PAT}}$}} & \multicolumn{2}{c|}{\textbf{Test$_{\text{IDEN}}$}} & \multicolumn{2}{c}{\textbf{Test$_{\text{ALL}}$}} \\
  \textbf{Training} & & \textbf{\textit{Acc.}} & \textbf{\textsc{Ori.} / \textsc{Adv.}} & \textbf{\textit{Acc.}} & \textbf{\textsc{Ori.} / \textsc{Adv.}} & \textbf{\textit{Acc.}} & \textbf{\textsc{Ori.} / \textsc{Adv.}} & \textbf{\textit{Acc.}} & \textbf{\textsc{Ori.} / \textsc{Adv.}} \\
  \midrule
  
  \multirow{7}{*}{\textbf{Train$_{\text{PRADA}}$}} & OSD (($\delta_{osd}=0.36$)) & 50.0 & 99.2 / 0.77 & 50.4 & 99.3 / 1.56 & 51.6 & 99.3 / 3.89 & 50.7 & 99.3 / 2.11 \\ 
  & PPL ($\delta_{ppl}=55.9$) & 76.0 & 82.1 / 69.8 & 57.7 & 84.0 / 31.5 & 50.4 & 83.6 / 17.2 & 61.1 & 83.2 / 39.0 \\
  & LA ($\delta_{la}=0.62$) & 81.1 & 79.0 / 83.2 & \textbf{80.5} & 78.8 / 82.1 & \textbf{58.1} & 77.5 / 38.8 & \textbf{73.0} & 78.4 / 67.5 \\
  & BERT-Base w/o query & 95.0 & 97.9 / 92.0 & 54.3 & 97.8 / 10.8 & 51.8 & 97.5 / 6.05 & 66.6 & 97.7 / 35.4 \\ 
  & BERT-Base w/ query & 95.2 & 97.8 / 92.6 & 59.5 & 97.7 / 21.4 & 56.1 & 97.3 / 14.9 & 69.9 & 97.6 / 42.2 \\
  & RoBERTa-Base w/o query & \textbf{99.8} & 99.9 / 99.8 & 50.1 & 99.9 / 0.39 & 49.9 & 99.8 / 0.00 & 66.1 & 99.9 / 32.4 \\
  & RoBERTa-Base w/ query & \textbf{99.8} & 99.9 / 99.8 & 50.1 & 99.9 / 0.39 & 49.9 & 99.7 / 0.06 & 66.1 & 99.8 / 32.4 \\
  \midrule

  \multirow{7}{*}{\textbf{Train$_{\text{PAT}}$}} & OSD (($\delta_{osd}=0.14$)) & 48.1 & 61.0 / 35.1 & 59.0 & 62.1 / 55.9 & \textbf{69.8} & 61.1 / 78.4 & 59.2 & 61.4 / 57.0 \\ 
  & PPL ($\delta_{ppl}=35.0$) & 72.0 & 62.0 / 82.1 & 59.9 & 61.7 / 58.0 & 51.4 & 61.8 / 41.0 & 60.9 & 61.9 / 59.9  \\
  & LA ($\delta_{la}=0.64$) & \textbf{81.2} & 77.5 / 84.9 & 80.6 & 77.0 / 84.2 & 58.4 & 76.0 / 40.8 & \textbf{73.1} & 76.8 / 69.5 \\
  & BERT-Base w/o query & 51.1 & 99.1 / 3.13 & 99.5 & 99.3 / 99.6 & 51.6 & 99.2 / 4.06 & 67.3 & 99.2 / 35.4 \\
  & BERT-Base w/ query & 50.3 & 99.7 / 0.95 & 99.7 & 99.7 / 99.6 & 57.4 & 99.8 / 15.0 & 69.1 & 99.7 / 38.5 \\
  & RoBERTa-Base w/o query & 55.6 & 99.7 / 11.6 & \textbf{99.9} & 99.9 / 99.9 & 50.0 & 99.9 / 0.17 & 68.4 & 99.8 / 36.9 \\
  & RoBERTa-Base w/ query & 55.1 & 99.8 / 10.4 & \textbf{99.9} & 99.8 / 99.9 & 50.1 & 99.9 / 0.20 & 68.2 & 99.8 / 36.6 \\
  \midrule

  \multirow{7}{*}{\textbf{Train$_{\text{IDEM}}$}} & OSD (($\delta_{osd}=0.16$)) & 48.9 & 71.9 / 25.8 & 58.3 & 71.9 / 44.7 & 69.8 & 70.7 / 68.9 & 59.2 & 71.5 / 46.9 \\ 
  & PPL ($\delta_{ppl}=24.3$) & 66.1 & 43.3 / 88.9 & 58.6 & 42.4 / 74.8  & 52.1 & 41.8 / 62.4 & 58.8 & 42.5 / 75.1 \\
  & LA ($\delta_{la}=0.78$) & \textbf{77.6} & 61.0 / 94.3 & \textbf{77.2} & 60.4 / 94.0 & 59.2 & 59.8 / 58.6 & 71.1 & 60.4 / 81.9 \\
  & BERT-Base w/o query & 51.8 & 95.1 / 8.45 & 60.8 & 95.0 / 26.6 & 95.6 & 95.6 / 95.7 & 69.9 & 95.2 / 44.6 \\
  & BERT-Base w/ query & 50.2 & 99.2 / 1.17 & 72.4 & 98.7 / 46.1 & 98.6 & 98.8 / 98.3 & \textbf{74.2} & 98.9 / 49.6 \\
  & RoBERTa-Base w/o query & 49.7 & 97.2 / 2.18 & 54.6 & 97.4 / 11.7 & 96.9 & 97.7 / 96.1 & 67.6 & 97.4 / 37.7 \\
  & RoBERTa-Base w/ query & 50.3 & 99.2 / 1.44 & 68.2 & 98.9 / 37.4 & \textbf{98.9} & 99.2 / 98.6 & 73.0 & 99.1 / 46.9 \\
  \midrule

  \multirow{7}{*}{\textbf{Train$_{\text{ALL}}$}} & OSD (($\delta_{osd}=0.16$)) & 48.9 & 71.9 / 25.8 & 58.3 & 71.9 / 44.7 & 69.8 & 70.7 / 68.9 & 59.2 & 71.5 / 46.9 \\ 
  & PPL ($\delta_{ppl}=44.0$) & 74.5 & 72.7 / 76.4 & 59.7 & 74.0 / 45.4 & 50.8 & 73.9 / 27.6  & 61.4 & 73.5 / 49.3 \\
  & LA ($\delta_{la}=0.67$) & 80.9 & 74.3 / 85.8 & 80.3 & 74.0 / 86.6 & 58.7 & 73.0 / 44.5 & 73.1 & 73.8 / 72.4 \\
  & BERT-Base w/o query & 93.1 & 95.6 / 90.7 & 97.6 & 95.9 / 99.3 & 94.2 & 96.0 / 92.5 & 95.0 & 95.8 / 94.2 \\
  & BERT-Base w/ query & 94.7 & 99.0 / 90.3 & 99.0 & 98.6 / 99.5 & 97.8 & 98.4 / 97.1 & 97.2 & 98.7 / 95.7 \\
  & RoBERTa-Base w/o query & 99.0 & 98.3 / 99.6 & 99.2 & 98.4 / 99.9 & 96.1 & 98.6 / 93.6 & 98.0 & 98.4 / 97.7 \\
  & RoBERTa-Base w/ query & \textbf{99.6} & 99.6 / 99.7 & \textbf{99.7} & 99.5 / 99.9 & \textbf{98.8} & 99.6 / 98.1 & \textbf{99.4} & 99.5 / 99.2 \\  
  
  \bottomrule
\end{tabular}
}
\caption{\label{tab.point.res}
The point-wise detection \textit{Accuracy} (\textit{Acc.}) of all baseline detectors. All detectors are trained or adjusted on the single-attack (e.g., Train$_{\text{PRADA}}$) or the whole-attack (i.e., Train$_{\text{ALL}}$) Train set, and evaluated on different Test set as summarized in Table~\ref{tab:data.summary}. Along with the detection results on all documents, the separate results on the original (\textsc{Ori.}) and adversarial (\textsc{Adv.}) documents are also reported. The best results in each group are marked in bold.}
\end{table*}
\begin{table*}[t]
\centering
\resizebox{\textwidth}{!}{
\begin{tabular}{ll|cc|cc|cc|c}
\toprule
  \multirow{2}{*}{\textbf{Detector}} & \multirow{2}{*}{\textbf{Thread} \boldmath $\delta$} & \multicolumn{7}{c}{\textbf{Test$_{\text{ALL}}$}} \\
  & & \textbf{MRR@10} & \textbf{MRR} & \textbf{\textit{Acc.}} & \textbf{\textsc{Ori.} / \textsc{Adv.}} & \textbf{\#DD} & \textbf{\textsc{Ori.} / \textsc{Adv.}} & \textbf{\#DR}~$\downarrow$ \\
  \midrule
  Original Ranking List & - & 0.3919 & 0.4000 & - & - & - & - & - \\
  Attacked Ranking List & - & \underline{0.3599} & \underline{0.3683} & - & - & - & - & - \\
  \midrule
  \multirow{2}{*}{OSD} & 0.16 & 0.1215 & 0.1240 & 69.6 &  69.7 / 46.9 & 7.93 & 7.61 / 0.31 & 0.74 \\
  & 0.66 & 0.3593 & 0.3677 & 99.7 & 99.9 / 0.00 & 0.00 & 0.00 / 0.00 & 0.00 \\
  \midrule
  
  \multirow{2}{*}{PPL} & 44.0 & 0.2734 & 0.2792 & 73.7 &  73.8 / 49.3 & 2.51 & 2.38 / 0.13 & 0.31 \\
  & 472.0 & 0.3599 & 0.3684 & 99.6 & 99.9 / 1.18 & 0.00 & 0.00 / 0.00 & 0.00 \\
  \midrule

  \multirow{2}{*}{LA} & 0.67 & 0.3040 & 0.3098 & 73.6 & 73.6 / 72.4 & 2.63 & 2.43 / 0.21 & 0.26 \\
  & 0.06 & 0.3599 & 0.3683 & 99.7 & 99.9 / 0.12 & 0.00 & 0.00 / 0.00 & 0.00 \\
  \midrule
  
  \multirow{2}{*}{BERT-Base w/o query} & 0.50 & 0.3793$^\uparrow$ & 0.3869$^\uparrow$ & 95.7 & 95.7 / 94.2 & 1.01 & 0.61 / 0.40 & 0.05 \\
   & 0.99 & \textbf{0.3856}$^\uparrow$ & \textbf{0.3936}$^\uparrow$ & 97.8 & 97.8 / 90.8 & 0.69 & 0.30 / 0.38 & 0.02 \\
  \midrule
  
  \multirow{2}{*}{BERT-Base w/ query} & 0.50 & 0.2803 & 0.2867 & 97.9 & 98.0 / 95.7 & 2.06 & 1.64 / 0.42 & 0.29 \\
   & 0.99 & 0.3345 & 0.3418 & 99.0 & 99.0 / 94.0 & 1.23 & 0.82 / 0.41 & 0.15 \\
  \midrule
  
  \multirow{2}{*}{RoBERTa-Base w/o query} & 0.50 & 0.3836$^\uparrow$ & 0.3916$^\uparrow$ & 98.4 & 98.4 / 97.7 & 0.65 & 0.25 / 0.40 & 0.02 \\
   & 0.99 & 0.3852$^\uparrow$ & 0.3932$^\uparrow$ & 98.8 & 98.8 / 96.8 & 0.57 & 0.17 / 0.40 & 0.02 \\
  \midrule
  
  \multirow{2}{*}{RoBERTa-Base w/ query} & 0.50 & 0.3447 & 0.3521 & 99.2 & 99.2 / 99.2 & 1.27 & 0.85 / 0.42 & 0.13 \\
   & 0.99 & 0.3561 & 0.3636 & 99.5 & 99.5 / 98.9 & 1.06 & 0.64 / 0.42 & 0.10 \\

  \bottomrule
\end{tabular}
}
\caption{\label{tab.list.res}
The list-wise detection results of all baseline detectors. The original or attacked ranking lists are produced by the victim model `msmarco-MiniLM-L-12-v2' under no or three adversarial documents for each query.
All detection methods are trained or adjusted on the whole-attack Train set (Train$_{\text{ALL}}$), and evaluated on the whole-attack Test set (Test$_{\text{ALL}}$) as summarized in Table~\ref{tab:data.summary}. 
\#DD denotes the average number of discarded documents ranked before the relevant document, and \#DR denotes the average number of discarded relevant documents.}
\end{table*}

\subsection{Experimental Results}
\subsubsection{Results of Point-wise Detection.}
The point-wise detection results are summarized in Table~\ref{tab.point.res}, which reflect the accuracy of the detection methods in classifying both the original and adversarial documents.

\textbf{The OSD-based and PPL-based detectors show limited effectiveness in identifying adversarial documents.} 
As seen in Table~\ref{tab.point.res}, regardless of whether the threshold is tuned on the single-attack set (e.g., Train$_{\rm PRADA}$) or the whole-attack set (i.e., Train$_{\rm ALL}$), both the OSD-based and PPL-based detectors achieves only about 60\% accuracy on the whole Test$_{\rm ALL}$ set.
As for PPL-based detector, this could be attributed to the significant overlap in perplexity values between the original and adversarial documents, making it challenging to find a suitable threshold for differentiation~\cite{DBLP:conf/ccs/LiuKTSSWLL22}.
Among the adversarial documents, IDEM-generated ones are comparatively more easily detected by the OSD-based detector, whereas PRADA-generated ones are relatively easier to be identified by the PPL-based detector.
As an example, when adjusting the OSD threshold ($\delta_{osd}=0.14$) on the Train$_{\rm PAT}$ set, the OSD-based detector achieve the highest detection accuracy on the Test$_{\rm IDEM}$ set, which suggests that the IDME attack method tends to add more query keywords into the original documents.
Similarly, when tuning the PPL threshold ($\delta_{ppl}=44.0$) on the whole-attack Train${\rm ALL}$ set, the detection accuracy is 74.5\% on the Test$_{\rm PRADA}$ set, but 59.7\% on the Test$_{\rm PAT}$ set and 50.8\% on the Test$_{\rm IDEM}$ set, which indicates that the PRADA attack method alters the perplexity of original documents a lot.

\textbf{The LA-based detector demonstrates greater reliability when faced with unknown adversarial documents.}
As shown in Table~\ref{tab.point.res}, when tuned on the Train$_{\rm PRADA}$ and Train$_{\rm PAt}$ sets, the LA-based detector achieves the best overall accuracy (about 73\%) on the Test$_{\rm ALL}$ set, and it even outperforms the supervised BERT and RoBERTa detectors, whose overall accuracy is less than 70\%.
This LA-based detector is trained only on an out-of-domain dataset in linguistics, yet it consistently achieves > 70\% accuracy on the Text$_{\rm ALL}$ set, indicating that the adversarial documents produced by existing attack methods (especially PRADA and PAT) do contain certain grammatical errors, which can be noticed by this LA model.
Additionally, the accuracy of LA-based detector on the Test$_{\rm IDEM}$ set (about 59\%) is significantly lower than that on other Test$_{\rm PRADA}$ and Test$_{\rm PAT}$ sets (about 80\%), which suggests that the IDEM's adversarial documents shall be more grammatically fluent and correct.

\textbf{The supervised BERT and RoBERTa detectors work effectively only when all types of adversarial documents are known.}
As indicated in Table~\ref{tab.point.res}, if the BERT and RoBERTa detectors are trained and tested on the same single-attack sets, such as from Train$_{\rm PRADA}$ to Test$_{\rm PRADA}$, they achieve accuracy above 95\%, but they fail to perform well (near to random guessing) when transferred to other single-attack sets, such as from Train$_{\rm PRADA}$ to Test$_{\rm IDEM}$.
An interesting exception is the transfer from Train$_{\rm IDEM}$ to Test$_{\rm PAT}$, where BERT and RoBERTa detectors achieve accuracy above 60\%, this might be due to the similarity in attack methods between IDEM and PAT, as well as the quality of IDEM's adversarial documents.
Furthermore, when the detectors are trained on Train$_{\rm ALL}$, which includes all types of adversarial documents, the BERT and RoBERTa models can successfully learn the distinctions between all adversarial documents and original documents, particularly the RoBERTa-base w/ query model.
When the query text is used to distinguish adversarial documents (i.e., w/ query), the detectors show improved performance, especially when the training data is Train$_{\rm IDEM}$ or Train$_{\rm ALL}$. 
However, even though the query text aids in point-wise detection, it can also negatively impact defense effectiveness in a ranking list due to its higher likelihood of misclassifying relevant documents.

\subsubsection{Results of List-wise Detection.}\label{sec.res.list}
The list-wise detection results are presented in Table~\ref{tab.list.res}, which further evaluate the effectiveness of detection defense on the quality of real text ranking.
During ranking attack, the ranking quality degrades when a low-ranked document is promoted to a higher rank than the relevant document.
On the other hand, during ranking defense, the ranking quality can be improved by discarding documents ranked before the relevant document while ensuring that the relevant document is not discarded.
To provide a clearer understanding of the results, we report two metrics in Table~\ref{tab.list.res}: the average number of filtered documents that ranked before the relevant document (\#DD), and the average number of relevant documents that are discarded (\#DR) for each test query, which provide insights into the impact of ranking defense on the improvement of ranking metrics.

\textbf{The OSD-based, PPL-based, and LA-based detectors do not offer any improvement.}
As seen in Table~\ref{tab.list.res}, when the candidate list contains at most three different adversarial documents (by PRADA, PAT and IDEM), the ranking performance of the victim NRM degrades from 0.39 to 0.36 in terms of MRR@10.
After utilizing the OSD-based, PPL-based and LA-based detectors to discard potential adversarial documents, using their thresholds in point-wise detection, the ranking quality deteriorates even further to 0.12, 0.27, and 0.30, respectively.
This degradation may occur due to the detectors wrongly discarding relevant documents. For instance, each query loses an average of 0.74, 0.31, and 0.26 relevant documents when using OSD-based, PPL-based, and LA-based detectors, respectively.
Moreover, even when tuning the thresholds of OSD-based, PPL-based, and LA-based detectors in terms of MRR@10, they can only preserve the ranking quality from deteriorating by avoiding discarding documents as many as possible.
For instance, the accuracy of these three detectors exceeds 99\%, indicating that they barely discard any documents to not further affect the attacked ranking quality.

\textbf{Supervised BERT and RoBERTa detectors without query text demonstrate the most significant recovery in ranking quality.}
As observed in Table~\ref{tab.list.res}, only two types of detectors, `BERT-Base w/o query' and `RoBERTa-Base w/o query', contribute to enhancing the ranking quality over the attacked ranking list, with a gain of 2-3 MRR points.
When comparing the detectors with and without query text, we notice that using query text (`w/o query') can improve the classification accuracy and increase the number of discarded documents (\#DD) contributing to the ranking metrics. However, it also leads to an increase in the number of discarded relevant documents (\#DR), indicating that the ranking list for some queries does not contain any relevant document.
Adversarial documents can convey pseudo-relevant information to mislead the victim NRMs, using this kind of text along with the query text for supervised classification training, the detection model tends to function more as a relevance scoring model, leading to relevant query-document pairs being more likely to be misclassified as adversarial and discarded.
Thus, the results suggest that using only the document text itself to train a classifier yields better defense for the ranking list. 
Similar to the unsupervised detectors like PPL, tuning the threshold (from 0.5 to 0.99) to make the supervised detectors more stringent in discarding documents can be helpful, as most of the documents in a ranking list are normal or original rather than adversarial.

\section{Related Work}
The robustness of neural IR models, such as neural ranking models (NRMs) and dense retrieval (DR) models, are increasingly attracting the attention of researchers.
Unlike the effectiveness which is about the average performance of a system, robustness cares more about the worst-case performance instead~\cite{DBLP:journals/tois/WuZGFC23}.
Existing studies have shed light on the robustness from various perspectives, such as performance variance~\cite{DBLP:journals/tois/WuZGFC23}, zero-shot domain transfer~\cite{DBLP:conf/acl/MaSA21}, query typos or variations~\cite{zhuang-zuccon-2021-dealing,DBLP:conf/ecir/PenhaCH22,DBLP:conf/ijcai/ChenLH0S22,DBLP:conf/sigir/ZhuangZ22,DBLP:conf/sigir/SidiropoulosK22}, document noises or errors~\cite{DBLP:conf/ecir/BazzoLVM20,DBLP:conf/cikm/AhmedB22,DBLP:journals/ipm/ChenHHSS23,DBLP:journals/jodl/OliveiraVACGRRM23}, and adversarial attacks~\cite{DBLP:journals/corr/abs-2008-02197,song-etal-2022-trattack,DBLP:journals/corr/abs-2204-01321,DBLP:conf/ccs/LiuKTSSWLL22,DBLP:conf/ictir/WangLA22,DBLP:conf/acl/ChenHY0S23,DBLP:conf/sigir/Liu0GR0FC23}.
These efforts are critical to know how would neural IR models behave in abnormal situations, before them enter into the real-world retrieval scenarios.

Adversarial ranking attacks add imperceptible perturbations into the target document to promote its position in the rankings, which can be regarded as a new type of black-hat search engine optimization (black-hat SEO;~\citealp{DBLP:journals/ftir/CastilloD10}).
Currently, several document manipulation methods have been proposed to craft the adversarial documents by word substitution~\cite{DBLP:journals/corr/abs-2204-01321,DBLP:conf/sigir/Liu0GR0FC23} or trigger injection~\cite{DBLP:conf/ccs/LiuKTSSWLL22,DBLP:conf/acl/ChenHY0S23,DBLP:conf/sigir/Liu0GR0FC23}, with respect to a specific query or a group of queries with the same topic~\cite{DBLP:conf/sigir/Liu0GR0FC23}.
In general, the focus has primarily been on designing attack methods, with limited research dedicated to defense strategies. However, it is essential to combine both attack and defense techniques to effectively advance the development of robust IR models. 
This study has taken a small step in this direction, aiming to encourage further exploration and the creation of adversarial ranking defense/detection methods, such as adversarial training~\cite{DBLP:conf/ecir/LupartC23} and certified robustness~\cite{DBLP:conf/cikm/WuZGCFRC22}, which play a vital role in promoting the development of robust real-world search engines.

\section{Conclusion}
In this study, we construct a benchmark dataset to support the study of adversarial ranking defense, and introduce two kinds of detection tasks in the point-wise and list-wise perspective.
Extensive experiments are carried out with several straightforward detection baselines, including unsupervised methods based on spamicity, perplexity, or linguistic acceptability, as well as supervised classifiers.
Experimental results offer valuable empirical insights that illuminate effective approaches for countering adversarial ranking attacks.

% \clearpage
\bibliography{custom}
\bibliographystyle{acl_natbib}

% \appendix

% \section{Example Appendix}
% \label{sec:appendix}

% This is an appendix.

\end{document}